\label{} 
\documentclass[prc,twocolumn,showpacs,floatfix,nofootinbib,preprintnumbers,superscriptaddress,amsmath,amssymb]{revtex4-1}

\usepackage{amsmath}           
\usepackage{amsfonts}
\usepackage{dcolumn}           
\usepackage{bm}
\usepackage[retainorgcmds]{IEEEtrantools}
\usepackage{color}
\usepackage{multirow}

\usepackage{graphicx,epsfig,color}

\definecolor{grey}{rgb}{0.93,0.93,0.93}
\definecolor{lred}{rgb}{1,0.4,0.4}

\newcommand{\be}{\begin{equation}}
\newcommand{\ee}{\end{equation}}
\newcommand{\bn}{\begin{eqnarray}}
\newcommand{\en}{\end{eqnarray}}
\newcommand{\ba}{\begin{array}}
\newcommand{\ea}{\end{array}}
\newcommand{\bc}{\begin{center}}
\newcommand{\ec}{\end{center}}
\newcommand{\bml}{\begin{mathletters}}
\newcommand{\eml}{\end{mathletters}}

\begin{document}


\title{Energy density functional for nuclei and neutron stars}

\author{J. Erler}
\affiliation{
Department of Physics and Astronomy, University of Tennessee, Knoxville, Tennessee 37996, USA
}%
\affiliation{
Physics Division, Oak Ridge National Laboratory, Oak Ridge, Tennessee 37831, USA
}%
\affiliation{
Division Biophysics of Macromolecules, German Cancer Research Center (DKFZ), Im Neuenheimer Feld 580, D-69120 Heidelberg, Germany
}%
\author{C.J. Horowitz}
\affiliation{
Department of Physics and CEEM, Indiana University, Bloomington, Indiana 47405, USA
}%
\affiliation{
Department of Physics and Astronomy, University of Tennessee, Knoxville, Tennessee 37996, USA
}%
\affiliation{
Physics Division, Oak Ridge National Laboratory, Oak Ridge, Tennessee 37831, USA
}%

\author{W. Nazarewicz}
\affiliation{
Department of Physics and Astronomy, University of Tennessee, Knoxville, Tennessee 37996, USA
}%
\affiliation{
Physics Division, Oak Ridge National Laboratory, Oak Ridge, Tennessee 37831, USA
}%
\affiliation{
Institute of Theoretical Physics, University of Warsaw, ul. Ho\.za 69,
PL-00-681 Warsaw, Poland 
}%

\author{M. Rafalski}

\affiliation{
Department of Physics and Astronomy, University of Tennessee, Knoxville, Tennessee 37996, USA
}%
\affiliation{
Physics Division, Oak Ridge National Laboratory, Oak Ridge, Tennessee 37831, USA
}%

\author{P. -G. Reinhard}
\affiliation{Institut fur Theoretische Physik II, Universitat Erlangen-Nurnberg,
Staudtstrasse 7, D-91058 Erlangen, Germany}

\date{\today}

\begin{abstract}
\begin{description}
\item[Background]
Recent observational data on  neutron star masses and radii
provide  stringent constraints on the  equation of state of neutron rich matter
\cite{LattRev}.
\item[Purpose]
We aim to develop a nuclear energy density functional that can be simultaneously applied to finite nuclei and neutron stars.
\item[Methods]
We  use the self-consistent nuclear density functional theory (DFT)  with Skyrme energy density functionals and covariance analysis to assess correlations between observables for finite nuclei and neutron stars. In a first step two energy functionals -- a high density energy functional giving reasonable neutron properties, and a low density functional fitted to nuclear properties -- are  matched. In a second step, we optimize a new functional using exactly the same protocol  as  in  earlier studies pertaining to nuclei but now including neutron star data. This allows direct comparisons of performance of the new functional relative to the standard one.
\item[Results] 
The new functional TOV-min  yields results for  nuclear bulk properties (energy, r.m.s. radius, diffraction radius, surface thickness) that are of the same quality as those  obtained with the established Skyrme functionals, including SV-min. When comparing SV-min and TOV-min, isoscalar
nuclear matter indicators  vary slightly while isovector properties are  changed considerably. 
We discuss  neutron skins, dipole polarizability, separation energies of the heaviest  elements, and proton and neutron drip lines. We confirm a correlation
between the neutron skin of $^{208}$Pb and the neutron star  radius.
\item[Conclusions]
We demonstrate that standard energy density functionals optimized to nuclear data  do not carry information on  the expected maximum neutron star mass, and that predictions can only be made within an extremely broad uncertainty band. 
For atomic nuclei, the new functional TOV-min performs at least as well as the standard nuclear functionals, but it also reproduces expected neutron star  data within assumed error bands.
This functional is expected to yield more reliable predictions in the region of very neutron-rich heavy nuclei.
\end{description}
\end{abstract}


\bigskip

\maketitle

\section{Introduction}
\label{sec.introduction}

We encounter the same material both in the laboratory and in outer space.  This basic principle has been remarkably powerful for centuries.  For Newton, the material was mass, and he observed that gravity acted in the same way on an apple and on the moon.  In the 19th century, the materials were the chemical elements.   Astrophysics grew from spectral lines being the same in the laboratory and in the cosmic space.  Indeed, the second element helium is named after the Greek God of the sun.   The material, we focus on in this paper, is neutron rich matter.   Both in the laboratory and in astrophysics, it has the same neutrons, the same strong interactions, and the same equation of state (EOS).   We aim to develop a universal nuclear energy functional that can be applied to stable, neutron rich, and very heavy nuclei, and to neutron rich matter in astrophysics including neutron stars.

A frequently adopted strategy is to optimize an energy density functional (EDF) to properties of finite nuclei.  However, this can leave many properties of bulk neutron matter poorly constrained.  Alternative optimization protocols are  based on considering both experimental data on finite nuclei and theoretical pseudo-data on  nuclear and neutron matter. For example, the 
SLy4 \cite{SLy4} EDF was fit to both nuclei and the variational Fermi hypernetted chain  calculations of neutron matter \cite{FHNC}.  Chamel {\it et al.} calculated the maximum mass of neutron stars for three EOS fit to finite nuclei and microscopic neutron matter calculations \cite{Chamel:2011aa}, and such analysis helped them to so select one EDF (BSk21) that does well on both kinds of data.  
However, it is to be noted that current microscopic calculations of neutron matter are highly model dependent; in particular, the impact of the poorly known three-neutron forces (especially their $T=3/2$ component) seems to be crucial \cite{Gan12a,Ste12a}.  Furthermore, four-nucleon forces \cite{4Nforces} may also be important, especially at high density.  Indeed, at this time, our ability to calculate neutron matter properties at high densities is fundamentally limited.  For example the chiral effective field theory calculations of Ref.~\cite{Heb10a} do not converge at densities much beyond normal nuclear (saturation) density.  Although microscopic approaches with phenomenological two- and three-nucleon interactions such as those of Ref.~\cite{Gan12a} do not explicitly depend on a chiral expansion, it is not clear how can they avoid large ambiguities that are present in a chiral approach at densities beyond where the chiral expansion converges.    

Recently, there has been considerable progress in determining neutron star (NS) masses and radii \cite{LattRev}.  Demorest {\it et al.} have accurately measured a 1.97 $\pm 0.04$ $M_\odot$ neutron star using Shapiro delay of the radio signal \cite{Dem10a}.  This model-independent result immediately provides an important lower limit for the maximum mass of a neutron star.  The EOS of neutron rich matter must have a high enough pressure to support {\it at least} this mass against collapse into a black hole.
In addition, neutron star radii and masses have been inferred from X-ray observations of both quiescent stars in globular clusters and stars undergoing thermonuclear bursts \cite{ozel,*Ozel2012,steiner}.  These results depend on atmosphere models and on the model assumed for the X-ray bursts \cite{Galloway2012}.  For example, Suleimanov {\it et al.} employ more sophisticated atmosphere models  \cite{Suleimanov2011,*Suleimanov2011a,*Suleimanov2012} and obtain a larger radius than Steiner {\it et al.} \cite{steiner},  while Ref. \cite{Servillat2012} finds that extracted neutron star radii depend on the assumed composition of the atmosphere.  In the future, one may be able to obtain neutron star radii from observations of X-ray burst oscillations \cite{Strohmayer}. 

In principle heavy ion collisions can provide data on the symmetry energy at high densities and help constrain energy functionals \cite{TsangRev}.  For example, Ref.~\cite{Tolos2012} studied EOS by means of kaon production at subthreshold energies.   However, the interpretation of present heavy ion measurements may be model dependent.  Finally, anticipated gravitational wave observations \cite{Owen2009} from neutron star mergers may provide constraints on EOS \cite{Bauswein2012}. 

Modern energy functionals such as UNEDF0 \cite{UNEDF0}, UNEDF1 \cite{UNEDF1}, and SV-min \cite{Klu09a} have been optimized to a large variety of nuclear data.  (For reviews of Skyrme Hartree-Fock (SHF) calculations of nuclei see for example Refs. \cite{Ben03aR,Erl11aR}.)  Ref. \cite{Sto07aR}  provides an example of an application of SHF approach to neutron star structure while Ref.~\cite{Dutra} also reviews  Skyrme EDFs and discuss their divergent predictions for neutron and nuclear matter.  (Note that some of the constraints imposed in their review may be model dependent.)  

Agrawal {\it et al.} \cite{agrawal} optimized a relativistic mean field (RMF) Lagrangian to nuclei and some assumed neutron matter properties.  Fattoyev {\it et al.} \cite{Fattoyev:2012km} discussed optimization of  isovector parameters of both RMF  and SHF models  using microscopic calculations of neutron matter.  Other RMF functionals  have also been determined while paying attention to some NS properties \cite{IUFSU,fattoyev}.  Low density matter at sub-saturation density has been considered in Ref.~\cite{Khan:2012ps}.  Finally, RMF  calculations at high density, coupled with virial expansion and nuclear statistical equilibrium calculations at low density,  have been used to generate  astrophysical EOS \cite{She11a,She11b}; these give the pressure as function of density, temperature, and proton fraction, and can be used in simulations of supernovae, neutron star mergers, and black hole formation.

We implicitly assume that neutron rich matter can be at least approximately described by a single EOS  from nuclear density to the central density of massive neutron stars. We do not exclude  phase transitions to non-nucleonic degrees of freedom (exotic components), but consider an EOS that changes smoothly with density. For instance,  a high density phase with hyperons has been considered in Refs.  \cite{Massot:2012pf,Chamel:2012ea}, and 
a hadron-quark crossover was considered in Refs.~\cite{Ozel10,Mas12}.
In general the appearance of exotic components could reduce the pressure of neutron rich matter.  However, the observation of a massive 1.97 $M_\odot$ neutron star \cite{Dem10a} strongly suggests that any softening of the EOS at high densities associated with phase transitions is modest.  Otherwise the EOS might not be able to support 1.97 $M_\odot$ against collapse to a black hole.  
Note that quark matter may have an  EOS similar to that of pure hadronic matter \cite{masquerade}.  In this case, NS mass and radius observations could not distinguish a quark-hadron hybrid star from a purely hadronic star.  In either case, the assumption of a single EDF should be applicable.

In this work,  we choose to optimize the nuclear EDF  to both nuclear data and neutron star masses and radii.  This has the advantage of treating finite nuclei and neutron star observables on an equal footing.  We consider a mixed set of fit-observables consisting of  varios bulk properties of spherical nuclei  and NS data, including the maximum mass and the radius of a 1.4 $M_\odot$ star.  A new functional is obtained by optimizing the coupling constants to  this extended dataset.  This approach allows one to see how the NS data impact various parts of the energy functional and predictions in the neutron-rich territory, especially around the neutron drip line.

Our paper is organized as follows. Section~\ref{sec.formalism} overviews our model. In Sec. \ref{subsec.existing} we describe our original SV-min functional that has been optimized to nuclear properties only.  Next in Sec. \ref{subsec.TOV} we review the TOV equations of general relativity that describe NS structure. In Sec. \ref{subsec.matching} we describe an intermediate procedure that simply matches a high density energy functional, that gives reasonable NS properties, to a low density functional adjusted to nuclear properties.   In Sec. \ref{subsec.fitTOV} we describe how we optimize EDF to both finite nuclei and NS properties simultaneously.   Results of this optimization are presented in Sec. \ref{sec.results}. The NS observables depend on properties of very neutron rich matter.  This should allow our functional fit to NS to make more accurate predictions for very neutron rich and for very heavy nuclei.  Therefore, in Sec. \ref{sec.drip} we apply our functional to nuclei near the drip lines over a very large range of mass number $A$ up to superheavy systems.  Finally, Sec. \ref{sec.conclusions} contains the conclusions of our work.

\section{Energy Density Functional Formalism}
\label{sec.formalism}

\subsection{Existing functionals fit to finite nuclei}
\label{subsec.existing}


This section briefly outlines the nuclear DFT in the self-consistent SHF
variant, and the optimization  strategy employed in this study. (For an
in-depth presentation, we refer the reader to
Refs.~\cite{Ben03aR,Erl11aR} and references cited therein). The main
ingredient of the SHF theory is an expression for the
total energy:
\begin{eqnarray}
E
&=& E_\mathrm{kin}  + \int \! d^3 r \; {\cal E}_\mathrm{Sk}
+ E_\mathrm{Coul}
  + E_\mathrm{pair}
  - E_\mathrm{corr},
\label{eq:Etot}
\end{eqnarray}
where $E_\mathrm{kin}$ is the kinetic energy;  ${\cal E}_\mathrm{Sk}$ -- the
Skyrme EDF; 
$E_\mathrm{Coul}$ -- the
Coulomb term (where the direct Hartree energy is treated exactly and
the exchange term in Slater approximation);
$E_\mathrm{pair}$ is the pairing functional;
and $E_\mathrm{corr}$ is the correlation term that accounts for dynamical
correlations (c.m. correction, rotational correction). The key piece
in the SHF approach is the Skyrme EDF, which represents
the effective nuclear interaction between nucleons and can be partly
derived by considering a low momentum expansion of the density matrix
\cite{Neg72a,Bog10aR,Cardob}: 
\begin{eqnarray}
{\cal E}_{\rm Sk}
& = & \sum_{T = 0,1} 
      C_T^{\rho}\left(1+D_T^{\rho}\,\rho^\alpha\right) \, \rho_{T}^{2}
      + C_T^{\Delta \rho} \, \rho_{T} \Delta \rho_{T}
\nonumber\\
   &&\qquad   
      + C_T^{\tau} \, \rho_{T} \tau_{T}
      + C_T^{J} \mathbb{J}^2_{T} 
      + C_T^{\nabla J} \rho_{T} \, \nabla\!\cdot\!\vec{J}_{T}
     \;,
\label{eq:basfunct}
\end{eqnarray}
where 
\begin{equation}
  \rho_{T=0}^{\mbox{}}
  =
  \rho_p+\rho_n
  \quad,\quad
  \rho_{T=1}^{\mbox{}}
  =
  \rho_p-\rho_n
\end{equation}
are the isoscalar and isovector densities, respectively.
The local densities
in the functional are the particle density $\rho_q$, kinetic-energy
density $\tau_q$ and spin-orbit density $\mathbb{J}_q$, where the isospin
index $q=p,n$ labels protons and neutrons (for details and definitions of the coupling constants,
see \cite{Ben03aR,Erl11aR,UNEDF0}).  The functional 
(\ref{eq:basfunct}) shows, in fact, only the time-even couplings which
suffice for calculations of ground states (g.s.)  of even-even nuclei.  The form   (\ref{eq:basfunct}) is used for
the parameterizations SV-min \cite{SVmin} and
TOV-min, which will be introduced here. We will also consider an
alternative to the standard density dependence ($\propto \rho^{2+\alpha}$)
in terms of Pade approximants yielding the parameterization RD-min
\cite{Erl10b}. 

Lacking still sufficiently precise input from ab-initio many-body
theories, we adjust the parameters of the Skyrme functional
phenomenologically. To that end, we have scrutinized  nuclei for
correlation effects
\cite{Klu08a} and chosen a large sample of spherical nuclei which
have negligible correlations thus being well described in pure
mean-field theory.
The compilation of fit-observables   used for the optimization of SV-min and 
TOV-min is given in Table~\ref{tab:fits}. 
\begin{table}[ht]
\caption{\label{tab:fits}
Compilation of phenomenological input for the optimization of SV-min \cite{Klu09a} and TOV-min EDFs.
Abbreviations used are: $BE$ -- g.s. binding energy;
$R_\mathrm{diff}$ -- charge diffraction radius;
$\sigma$ -- charge surface thickness;
$r_\mathrm{ch}$ -- charge r.m.s. radius;
$\epsilon_{ls}$ -- spin-orbit splitting of selected
single-particle states;
NS --  NS data (only for TOV-min).
}
\begin{ruledtabular}
\begin{tabular}{l|l}
$BE$: & $^{36-52}$Ca, $^{68}$Ni, $^{100,126-134}$Sn,  $^{204-214}$Pb,
       $^{34}$Si, $^{36}$S, \\
  &     $^{38}$Ar, $^{50}$Ti,
       $^{86}$Kr, $^{88}$Sr, $^{90}$Zr
 $^{92}$Mo, $^{94}$Ru,  $^{96}$Pd, $^{98}$Cd,\\
&  
       $^{134}$Te, $^{136}$Xe, $^{138}$Ba, $^{140}$Ce, $^{142}$Nd, 
       $^{144}$Sm, $^{146}$Gd, 
\\
    &  $^{148}$Dy,
       $^{150}$Er, $^{152}$Yb, 
      $^{206}$Hg, $^{210}$Po, $^{212}$Rn, $^{214}$Ra, \\
    &  
      $^{216}$Th, $^{218}$U
\\[2pt]
$R_\mathrm{diff}$:
     & $^{16}$O, $^{40-44,48}$Ca, $^{58-64}$Ni, $^{118-124}$Sn,
       $^{204-208}$Pb,
       $^{50}$Ti,
\\
  &
       $^{52}$Cr, $^{54}$Fe, $^{86}$Kr,  $^{88}$Sr, $^{90}$Zr, $^{92}$Mo, $^{138}$Ba, $^{142}$Nd 
\\[2pt]
$\sigma$: & $^{16}$O, $^{40-44,48}$Ca, $^{60-64}$Ni, $^{118,122-124}$Sn,
       $^{204-208}$Pb, \\
&       $^{50}$Ti, $^{86}$Kr,
       $^{88}$Sr, $^{90}$Zr,
       $^{92}$Mo, $^{138}$Ba, $^{142}$Nd 
\\[2pt]
$r_\mathrm{ch}$: 
    &  $^{16}$O, $^{40-48}$Ca, $^{108,118-124}$Sn,  $^{200-214}$Pb,
       $^{36}$S, $^{38}$Ar, \\
    &   $^{50}$Ti, ${52}$Cr, $^{54}$Fe,
       $^{86}$Kr, $^{88}$Sr,  $^{90}$Zr, $^{92}$Mo,
\\
    &  
       $^{136}$Xe, $^{138}$Ba, $^{140}$Ce, $^{142}$Nd, 
       $^{144}$Sm, $^{146}$Gd, $^{148}$Dy,
\\
    & $^{150}$Er,
      $^{206}$Hg, $^{210}$Po, $^{212}$Rn, $^{214}$Ra 
\\[2pt]
$\epsilon_{ls}$:  & $^{16}$O($1p_n,1p_p$) $^{132}$Sn($2p_p,2d_n$), 
     $^{208}$Pb($2d_p,1f_n,3p_n$) 
\\[2pt]
NS: & $M_{\mathrm{max}}$, R($1.4 M_{\odot}$)
\\[2pt]
other: & $\delta r^2(^{214-208}\mathrm{Pb})$, odd-even mass differences 
\end{tabular}
\end{ruledtabular}
\end{table}

The model parameters are
adjusted to these data by virtue of a least-squares fit, for details
see Ref.~\cite{SVmin}. The unconstrained fit of the functional
(\ref{eq:basfunct}) to the  dataset from  Table~\ref{tab:fits} yields the
parameterization SV-min. Note that only information from finite nuclei
enters its calibration. A similar fit using the same data but a
modified density dependence yielded EDF RD-min \cite{Erl10b}. 

\begin{figure}
\includegraphics[width=0.9\columnwidth]{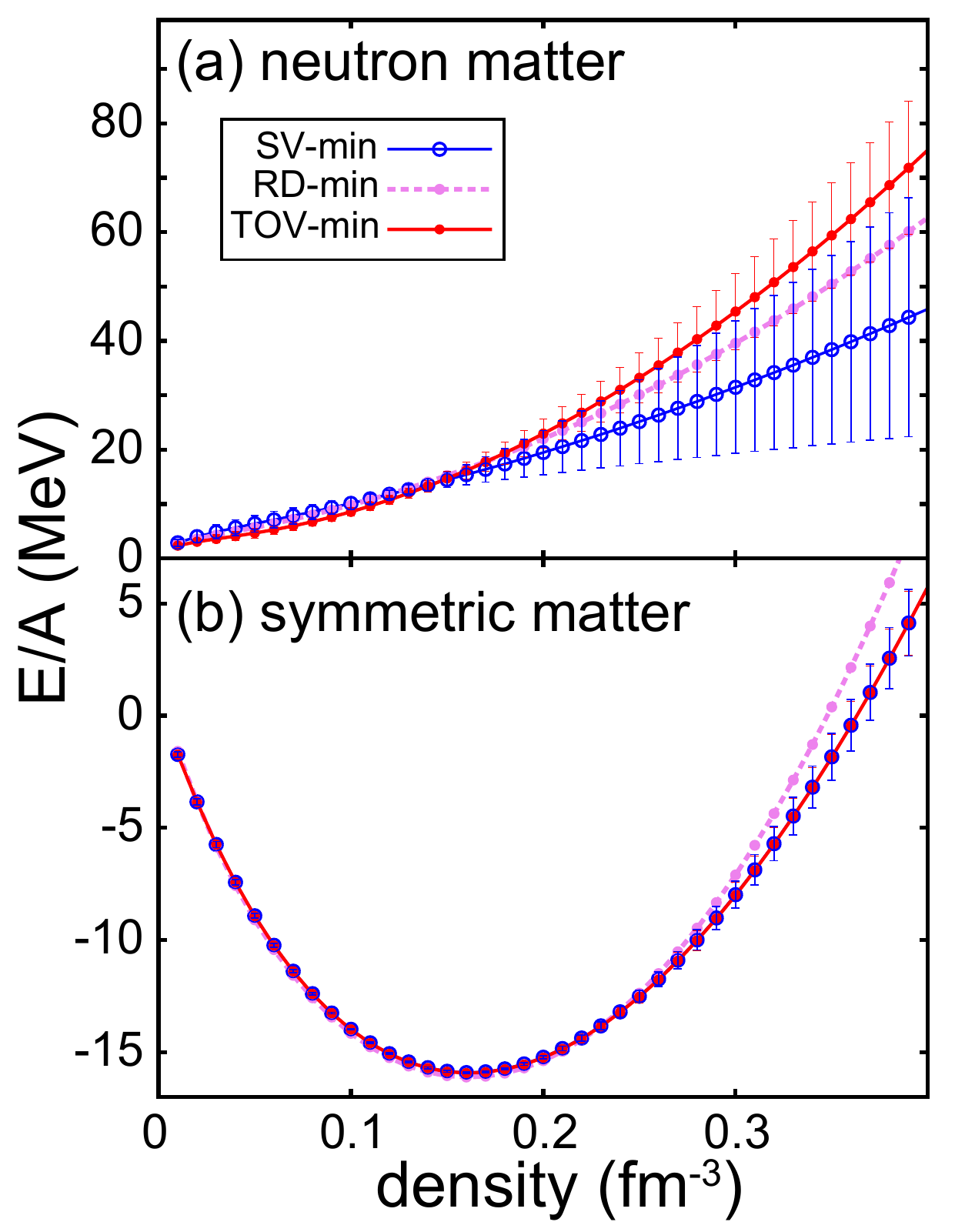}
\caption{\label{fig:EoS}
(Color online) Binding energy per nucleon in homogeneous  neutron matter (a) and symmetric matter (b)  as a function of density.
Shown are results for the parameterizations SV-min \cite{Klu09a}, RD-min \cite{Erl10b}, and TOV-min (this work).
For SV-min and TOV-min, we show also error bars representing the uncertainty in the extrapolation. The error bars for
RD-min are comparable to those of SV-min and dropped for better readability of the figure.}
\end{figure}
The well developed technique of least-squares covariance analysis yields not only
optimized parameterizations which can be used to extrapolate to
unknown regions, but it also allows to estimate  an extrapolation
error. This is illustrated in Fig.~\ref{fig:EoS}, which shows
the binding energy curves in homogeneous nuclear matter and neutron
matter for SV-min, RD-min, and TOV-min together with their
extrapolation uncertainties. Note that the uncertainty is increasing
dramatically with increasing density for SV-min, which was
calibrated to finite nuclei only (cf. Ref. \cite{Naz10a} for more discussion).  As the neutron matter EOS is
the main ingredient for the description of neutron stars, predictions
for neutron stars with standard Skyrme EDFs (here SV-min and RD-min)
are bound to be plagued by large uncertainties, see discussion below.

\subsection{TOV equations and neutron star structure}
\label{subsec.TOV}

The structure of a non-rotating cold neutron star can be characterized by the relationship between the gravitational mass $M$ and the radius $R$ of the star
\cite{LattRev}. This relation is obtained by solving the Tolman-Oppenheimer-Volkov (TOV) equations \cite{Opp39,Tol39}:
\begin{eqnarray}
\frac{d p(r)}{dr} &=& -\frac{G\epsilon(r) M(r)}{c^2 r^2}
\left[ 1+\frac{p(r)}{\epsilon(r)}\right] \label{eq:TOV1}\\
&&\left[1 + \frac{4 \pi r^3p(r)}{c^2 M(r)}\right]
\left[1-\frac{2G M(r)}{c^2r}\right]^{-1}, \nonumber\\
M(r) &=& 4\pi \int^r_0 r'^2 dr' \rho(r')=4 \pi \int_0^r r'^2 dr' \epsilon(r')/c^2. 
\label{eq:TOV2}  
\end{eqnarray}
Here $\rho(r)=\epsilon/c^2$ is the mass density at the distance $r$  and $\epsilon(r)$ is the corresponding energy density.  The enclosed mass $M(r)$ is the mass inside a sphere of radius $r$, and $p$ is the pressure.

The main physics ingredient necessary to solve equations (\ref{eq:TOV1}) and (\ref{eq:TOV2}) is the equation of state (EOS) providing the pressure as a function of the energy density $p=p(\epsilon)$. The coupled equations for $p(r)$ and $M(r)$ can be integrated starting from $r = 0$, and a value for the pressure at the center of the star $p_0$, up to the point where $p(R) = 0$. The result is the mass radius relation $M = M(R)$, which is expected to be consistent with our assumptions of a  maximal mass of $2.2M_{\odot}$ and a radius of $12.5$km for a $1.4 M_{\odot}$ neutron star, see below.
As a neutron star is not only made of neutrons, but is a mixture of neutrons, protons, electrons, and possibly muons -- due the weak decay of neutrons and electron capture processes on protons (for higher pressure even hyperons or kaons) -- the EOS has to be calculated for $\beta$-equilibrium \cite{Sto07aR}.
Neutron star matter is characterized by the following processes:
\begin{equation}
n \longleftrightarrow p + e^{-} \longleftrightarrow  p + \mu^{-}. 
\end{equation}
The corresponding chemical potentials should fulfill  the  $\beta$-equilibrium conditions:
\begin{equation}
\mu_n = \mu_p + \mu_e ,\quad \mu_\mu = \mu_e.
\end{equation}
Here each chemical potential is given by the canonical relation:
\begin{equation}
\mu_j = \frac{\partial \epsilon}{\partial n_j},
\end{equation}
with the total energy density $\epsilon$ containing all nucleon and lepton contributions for the particle number densities $n_j$ (for details see \cite{Sto07aR}). In addition, charge neutrality has to be fulfilled:
\begin{equation}
n_p=n_e+n_\mu.
\end{equation}
Lepton chemical potentials are calculated for noninteracting Fermi gases. In the following, all EOS are calculated at $\beta$-equilibrium.
\begin{figure}[ht]
\includegraphics[width=0.9\columnwidth,clip]{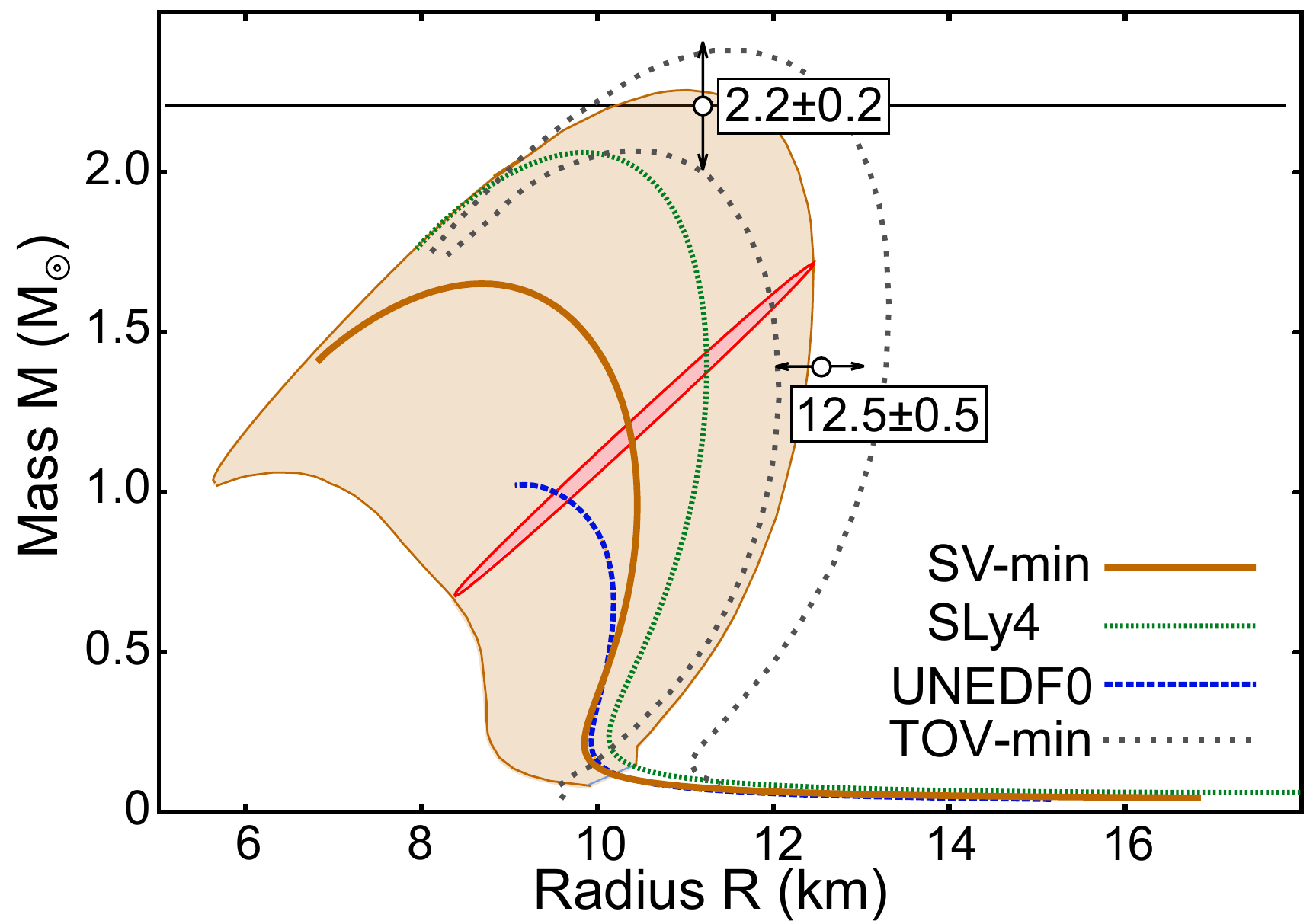}
\caption{
(Color online) Mass-radius relation of SLy4 \cite{SLy4}, UNEDF0 \cite{UNEDF0} and SV-min \cite{Klu09a}. The uncertainty band for SV-min is shown. This band is estimated by calculating the covariance ellipsoid for the mass $M$ and the radius $R$ at each point of the SV-min curve 
as indicated by the ellipsoid. Also depicted (dotted lines) are uncertainty limits for TOV-min.
}
\label{fig:uncert-SVmin}
\end{figure}

The results for the mass-radius relation of neutron stars $M(R)$ is shown in Fig. \ref{fig:uncert-SVmin} for  SV-min, SLy4,  and UNEDF0. In addition, the estimated uncertainty band for a prediction using SV-min is shown. As both observables are obviously correlated it is not possible to estimate the uncertainty for mass $M$ and radius $R$ separately. For this reason, the error band is obtained by calculating the covariance ellipsoid for mass $M$ and radius $R$ of a neutron star for each point of the $M(R)$ curve as indicated by the ellipsoid in Fig. \ref{fig:uncert-SVmin}. This ellipsoid is a measure for the correlation of two observables as presented in \cite{Naz10a}, and provides information about the uncertainty of an observable in dependence of a second one. The area covered by all ellipsoids can be interpreted as the error band for a prediction using SV-min.

We now discuss reasonable expected values for the maximum neutron star mass $M_{\rm max}$ and the radius of a $1.4 M_\odot$ neutron star $R_{1.4}$.  Demorest {\it et al.} have observed a $1.97\pm 0.04\ M_\odot$ star \cite{Dem10a}, so clearly $M_{\rm max}> 1.97 M_\odot$.  However elementary population synthesis considerations imply that the observation of a $1.97M_\odot$ star suggest that the actual maximum mass must be, at least, somewhat greater than this value.   This will allow for a reasonable probability to observe a $1.97 M_\odot$ star, if one draws neutron star masses from a realistic distribution.  Therefore we assume that
\begin{equation}
M_{\rm max}=2.2\pm 0.2\ M_\odot\, .
\label{eq:Mmax}
\end{equation}
This choice for $M_{\rm max}$ also provides room to accommodate the observation of somewhat more massive stars in the future.   

X-ray observations can provide information on neutron star radii.  For example, from luminosity and surface temperature measurements one can infer an emitting area.  However, the interpretation of X-ray observations may be model dependent.  One can be sensitive to the employed neutron star atmosphere model and or the assumed model for photosphere radius expansion X-ray bursts.  The radius of a 1.4 $M_\odot$ star $R_{1.4}$  between 10.4 and 12.9 km has been inferred in Ref.~\cite{Steiner2012}.  However Refs.~\cite{Suleimanov2011,*Suleimanov2011a,*Suleimanov2012} use more sophisticated neutron star atmospheres to model a long X-ray burst and infer larger radii.  At this point, we believe the final word on X-ray observations of neutron star radii has not yet been written.  Future observations and/or more sophisticated theoretical interpretations may change the inferred radii.  Therefore, at this time, we somewhat arbitrarily adopt a value for $R_{1.4}$ near the upper end of  range of Ref.~\cite{Steiner2012}:
\begin{equation}
R_{1.4}=12.5 \pm 0.5\ {\rm km}\, .
\label{eq:R1.4}
\end{equation}   
Here, the assumed 0.5 km error is somewhat arbitrary.  This value insures that fits to both nuclear and NS data give reasonable weight to the neutron star observations, see below.

As shown in Fig. \ref{fig:uncert-SVmin} the established Skyrme functionals can not reproduce the expected $M_{\rm max}$ (\ref{eq:Mmax}) and $R_{1.4}$ (\ref{eq:R1.4}).  However, these values are at the edges of the error band, and they can be reached by a new EDF fit  that includes NS data. 
Note that we do not explicitly include an EOS for the nonuniform neutron star crust.  Instead we simply use the EOS for uniform matter implied by a given Skyrme interaction even at low density.  This error has essentially no impact on $M_{\rm max}$.  However,  it somewhat reduces the radius of $1.4 M_\odot$ stars compared to predictions that employ realistic crust EOS at low densities.  Furthermore, this approximation significantly underestimates the radius of very low mass stars.

\subsection{Matching to a high density equation of state}
\label{subsec.matching}

\begin{figure}[h]
\includegraphics[width=0.9\columnwidth]{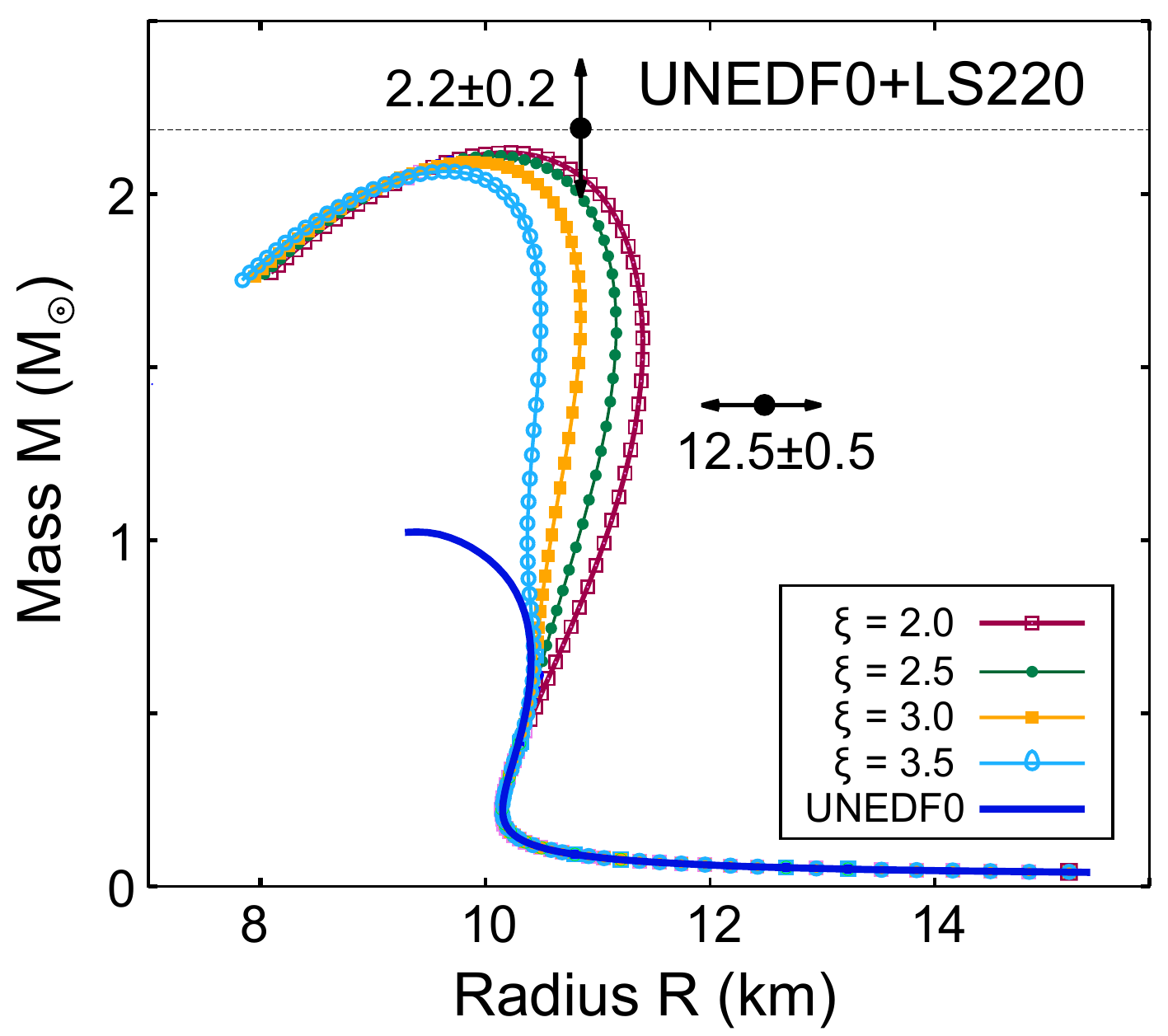}
\caption{(Color online) 
The mass-radius relation of an EOS matching the EDF UNEDF0 at low densities with the functional LS220 at high densities for matching densities $\rho = \xi\rho_0$ with $\xi =2.0$, 2.5, 3.0, and 3.5 (\ref{eq:switch}).
}
\label{fig-LS220_ksi}
\end{figure}

As a first step in creating a Skyrme functional for predicting properties of both finite nuclei and NS data,  we match an established nuclear Skyrme functional to an ansatz for the equation of state at high densities, for which we take  LS220 \cite{Lat91a}. To deal with neutron stars, it is necessary to consider a wide range of densities, starting from below the  nuclear saturation density $\rho_0 \approx 0.16\,\textrm{fm}^{-3}$ up to as high as $\rho \sim 10\,\rho_0$. The question is what is the optimal density to switch between the two equations of state. To answer this question, the following ansatz for a combined equation of state can be made
\begin{eqnarray}
P_{\textrm{c}}(\rho) 
&=& f(\rho)P_{\textrm{EDF}}(\rho) + (1-f(\rho))P_{\textrm{high}}(\rho),
\end{eqnarray}
\begin{eqnarray}
 \left(\frac{E}{A}\right)_{\textrm{c}}(\rho)
&=& \int_{0}^{\rho} \frac{P_{\textrm{c}}(\rho')}{\rho'^2} d\rho', 
\end{eqnarray}
where a switch factor
\begin{equation}
f(\rho) = \left(1+e^{\frac{\rho-\xi\rho_0}{a_{\rho}}}\right)^{-1}
\label{eq:switch}
\end{equation}
ensures a smooth matching. Here $P_{\textrm{EDF}}(\rho)$ is the pressure for a Skyrme energy functional and $P_{\textrm{high}}(\rho)$ is the pressure for the assumed high density functional. In the switch factor, $a_\rho=0.2\,\rho_0$ is a fixed diffuseness parameter and $\xi$ is a parameter that determines the matching density.  The advantage of this matching procedure is that any energy functional can be used to describe finite nuclei, while still providing a reasonable description of neutron stars.  Figure \ref{fig-LS220_ksi} shows the neutron star mass versus radius relation from combining the UNEDF0 and LS220 functionals, with several values of $\xi$. The maximum mass is seen to be relatively insensitive to $\xi$. Based on this result one can conclude that   
the low-density, or nuclear,  part of the EOS carries no information  on $M_{\rm max}$. That is, it makes little sense to scrutinize existing nuclear EDFs  with respect to this quantity. On the other hand, the radius of a $1.4\,M_{\odot}$ neutron star can vary by up to 1 km depending on the transition density used to match the two equations of state. 

\subsection{Optimizing EDF to nuclear and  NS data}
\label{subsec.fitTOV}

\begin{figure}[htb]
\includegraphics[width=0.8\columnwidth]{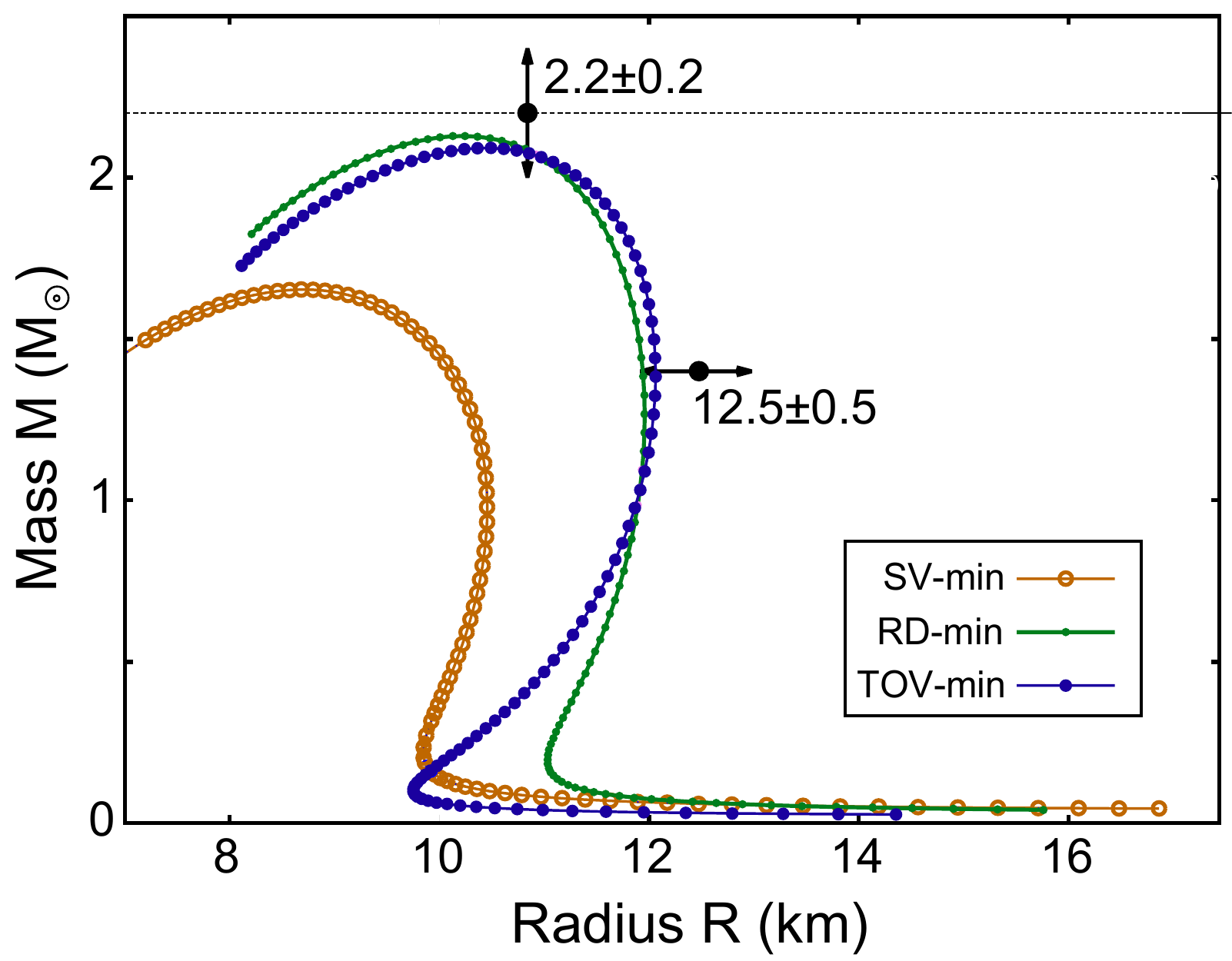}
\caption{\label{fig:MR}
(Color online) The mass radius relation for neutron stars obtained with SV-min, RD-min, and TOV-min.}
\end{figure}

In this section we describe optimization of a single EDF to both nuclear data listed in Table~\ref{tab:fits} and to the maximum mass $M_{\rm max}$ and radius $R_{1.4}$ of neutron stars, as indicated in Fig.~\ref{fig:uncert-SVmin}.  We use exactly the same optimization protocol as in Ref.~\cite{Klu09a}.  As a starting point, we choose the Skyrme EDF SV-min.  To include the additional information for neutron stars, the TOV equations (\ref{eq:TOV1}) and (\ref{eq:TOV2}) are solved at each step of the optimization procedure, assuming matter in $\beta$-equilibrium.  The result is the  new  Skyrme EDF TOV-min (see Table \ref{tab:params}). 
\begin{table}[ht]
\caption{\label{tab:params}
Optimized parameter set of TOV-min expressed in terms of the traditional
$(t, x)$ parameterization of the Skyrme force and in terms of $C$ coupling constants.
}
\begin{ruledtabular}
\begin{tabular}{l|r|l|r}
$t_0$ (MeV$\cdot$fm$^3$)  		& -2129.735   & $C^{\rho\rho}_{00}$ (MeV$\cdot$fm$^3$)		&   -798.650     \\[0.13cm]
$t_1$ (MeV$\cdot$fm$^5$)  		&   305.398   & $C^{\rho\rho}_{10}$ (MeV$\cdot$fm$^3$)		&    356.703     \\[0.13cm]
$t_2$ (MeV$\cdot$fm$^5$)  		&  362.532    & $C^{\rho\rho}_{0D}$ (MeV$\cdot$fm$^{3+3\alpha}$)&    872.316     \\[0.13cm]
$t_3$ (MeV$\cdot$fm$^{3+3\alpha}$)      &   13957.064 & $C^{\rho\rho}_{1D}$ (MeV$\cdot$fm$^{3+3\alpha}$)&   -524.922     \\[0.13cm]
$x_0$  					& 0.169949    & $C^{\rho\tau}_{0}$ (MeV$\cdot$fm$^5$) 		&    9.02926     \\[0.13cm]
$x_1$  					& -3.399480   & $C^{\rho\tau}_{1}$ (MeV$\cdot$fm$^5$) 		&    52.5831     \\[0.13cm]
$x_2$  					& -1.782177   & $C^{\rho\Delta\rho}_{0}$ (MeV$\cdot$fm$^5$) 	&   -55.0048     \\[0.13cm]
$x_3$  					&  0.402634   & $C^{\rho\Delta\rho}_{1}$ (MeV$\cdot$fm$^5$) 	&   -97.5411     \\[0.13cm]
$b_4$ (MeV$\cdot$fm$^5$)  		& 37.07357    & $C^{\rho\nabla J}_{0}$ (MeV$\cdot$fm$^5$)   	&   -79.1223     \\[0.13cm]
$b'_4$ (MeV$\cdot$fm$^5$) 		& 84.09737    & $C^{\rho\nabla J}_{1}$ (MeV$\cdot$fm$^5$)  	&   -42.0487     \\[0.13cm]
\hline\\[-0.2cm]
$\alpha$ & \multicolumn{3}{c}{0.250388065}        \\[0.13cm]
$V_\mathrm{pair,p}$ (MeV$\cdot$fm$^{6}$)  & \multicolumn{3}{c}{630.516}  \\  [0.13cm]
$V_\mathrm{pair,n}$ (MeV$\cdot$fm$^{6}$)  & \multicolumn{3}{c}{629.072}  \\[0.13cm]
$\rho_{0,\mathrm{pair}}$ (fm$^{-3}$)  & \multicolumn{3}{c}{0.201322348} \\  [0.13cm]
$\frac{\hbar^2}{2m_p}$ (MeV$\cdot$fm$^2$)& \multicolumn{3}{c}{20.7498207} \\[0.13cm]
$\frac{\hbar^2}{2m_n}$ (MeV$\cdot$fm$^2$)& \multicolumn{3}{c}{20.7212601} 
\end{tabular}
\end{ruledtabular}
\end{table}
Figure~\ref{fig:MR} presents the mass-radius relation for TOV-min together with results for SV-min  and RD-min.  The new functional performs well in reproducing the expected maximum mass and radius within the adopted errors (indicated by arrows).  Note also that RD-min, using the modified density dependence, also fits NS data without considering NS information during the optimization process.


\section{TOV-min performance}
\label{sec.results}

In this section results for the TOV-min EDF are presented and compared to results for some existing Skyrme functionals.

\subsection{TOV-min properties} 

To study the performance of the new functional, Table \ref{tableone} lists the root mean square errors of SV-min, RD-min, and TOV-min on   nuclear fit-observables: g.s. binding energy $BE$, charge diffraction radius $R_{\rm diff}$, surface thickness $\sigma$ and r.m.s. charge radius $r_{\rm ch}$, and pairing gaps for protons and neutrons ($\Delta_p$ and $\Delta_n$).
\begin{table}[htb]
\caption{\label{tableone}
Root mean square errors on nuclear fit-observables obtained by the parameterizations SV-min, RD-min, and the new functional TOV-min.
}
\begin{ruledtabular}
\begin{tabular}{l|lll} 
  \multicolumn{1}{c}{r.m.s. error}  
  & \multicolumn{1}{c}{SV-min}  
  & \multicolumn{1}{c}{RD-min}
 & \multicolumn{1}{c}{TOV-min} 
\\
\hline\\[-0.2cm]
$BE$ (MeV)   &  0.62  &  0.63 & 0.66 \\
$R_{\rm diff}$ (fm)     &  0.028 &  0.030& 0.028  \\
$\sigma$ (fm) &  0.022 &  0.022& 0.022  \\
 $r_{\rm ch}$   (fm)    &  0.014 &  0.013& 0.014 \\
$\Delta_p$ (MeV) &  0.11 &  0.11& 0.11 \\
$\Delta_n$ (MeV) &  0.14 &  0.14& 0.14\\[3pt]
Total {$\chi^2$} & 53.22 & 55.90  & 54.95
\end{tabular}
\end{ruledtabular}
\end{table}
Adding NS data has only a minor impact on the binding energy, while there is no change in performance for other nuclear bulk properties. This results in only a small increase in the total $\chi^2$ and shows that including NS data has almost no influence on nuclear bulk properties considered. 

\begin{table}[htb]
\caption{\label{tabletwo}
Nuclear matter properties as defined in \cite{Klu09a} for SV-min, TOV-min,  RD-min, and UNEDF0. For the isovector properties $\kappa$, $a_\mathrm{sym}$, and $L_\mathrm{sym}$ the calculated uncertainties are indicated.  The surface energy coefficient $a_{\rm surf}$ and surface-symmetry energy coefficient $a_{\rm ssym}$ are also given.}
\begin{ruledtabular}
\begin{tabular}{l|cccc}
 & SV-min & TOV-min & RD-min & UNEDF0 \\
\hline\\[-0.2cm]
{$\rho_0$ (fm$^{-3}$)} & 0.1610 &0.1610 &0.1611 & 0.1605  \\
{$E/A$ (MeV)} &-15.91 &-15.93 &-16.11 & -16.06    \\
{$K$ (MeV)} & 222 & 222& 231 & 230 \\
{$m^*/m$} & 0.95 & 0.94 &0.90 & 0.90 \\
{$a_\mathrm{sym}$ (MeV)} & 30.7$\pm$1.9 & 32.3$\pm$1.3 &32.1$\pm$2.1 & 30.5 \\
{$\kappa$}  &0.08$\pm$0.29 &0.21$\pm$0.26 &0.04$\pm$0.32 & 0.25\\  
{$L_{\mathrm{sym}}$ (MeV)} & 45$\pm$26 &76$\pm$15 &60$\pm$32 & 45 \\ 
$a_{\rm surf}$ (MeV)&17.6 & 17.6 & 17.6 &18.7 \\
$a_{\rm ssym}$ (MeV) &-51 & -44  &-55 & -44 
\end{tabular}
\end{ruledtabular}
\end{table}
Table \ref{tabletwo} shows nuclear matter properties as defined in 
Ref.~\cite{Klu09a,UNEDF0}.  In addition, the surface energy coefficient $a_{\rm surf}$ and surface-symmetry energy coefficient $a_{\rm ssym}$ are shown.  These are calculated from a leptodermous expansion of the energy functional (for details see Ref.~\cite{Rei06}).  Results for isoscalar properties ($\rho_\mathrm{eq}$, $E/A$, $K$, $m^*/m$ and $a_{\rm surf}$ ) vary slightly between SV-min and TOV-min, while results for isovector properties ($a_\mathrm{sym}$, $\kappa$, $L_\mathrm{sym}$ and $a_{\rm ssym}$) differ considerably as a result of neutron star constraints.
Of particular interest in the context of neutron structure are: the symmetry
energy $S_2$ at saturation density, $a_\mathrm{sym}=S_2(\rho_0)$, and 
the slope of the symmetry energy 
\begin{equation}\label{Lsym}
\left.  L_{\mathrm{sym}}=3\rho_0\frac{dS_2}{d\rho}\right|_{\rho_0}.
\end{equation}
The values of $a_{\rm surf}$ and $L_{\mathrm{sym}}$ in Table~\ref{tabletwo}
are consistent (within error bars) with the experimentally allowed  range
\cite{TsangRev, LattRev,LasttimerLim2012}.

\subsection{Correlations, neutron skin and dipole polarizability \label{cha:corr}}

A quantity characterizing the correlation between two observables $A$ and $B$ within a given model is the Pearson product-moment correlation coefficient $C_{AB}$ \cite{Bra97aB}, where a value $C_{AB} = 1$ means the observables are fully correlated and $C_{AB} = 0$ implies they are totally uncorrelated.  
The covariance analysis in terms of $C_{AB}$ has been proven useful in previous surveys \cite{Naz10a,FatPie11,Piek12,Fattoyev:2012km,ReinNaz12,Fatt12}.
\begin{figure}[htb]
\centerline{
{\includegraphics[width=0.9\columnwidth]{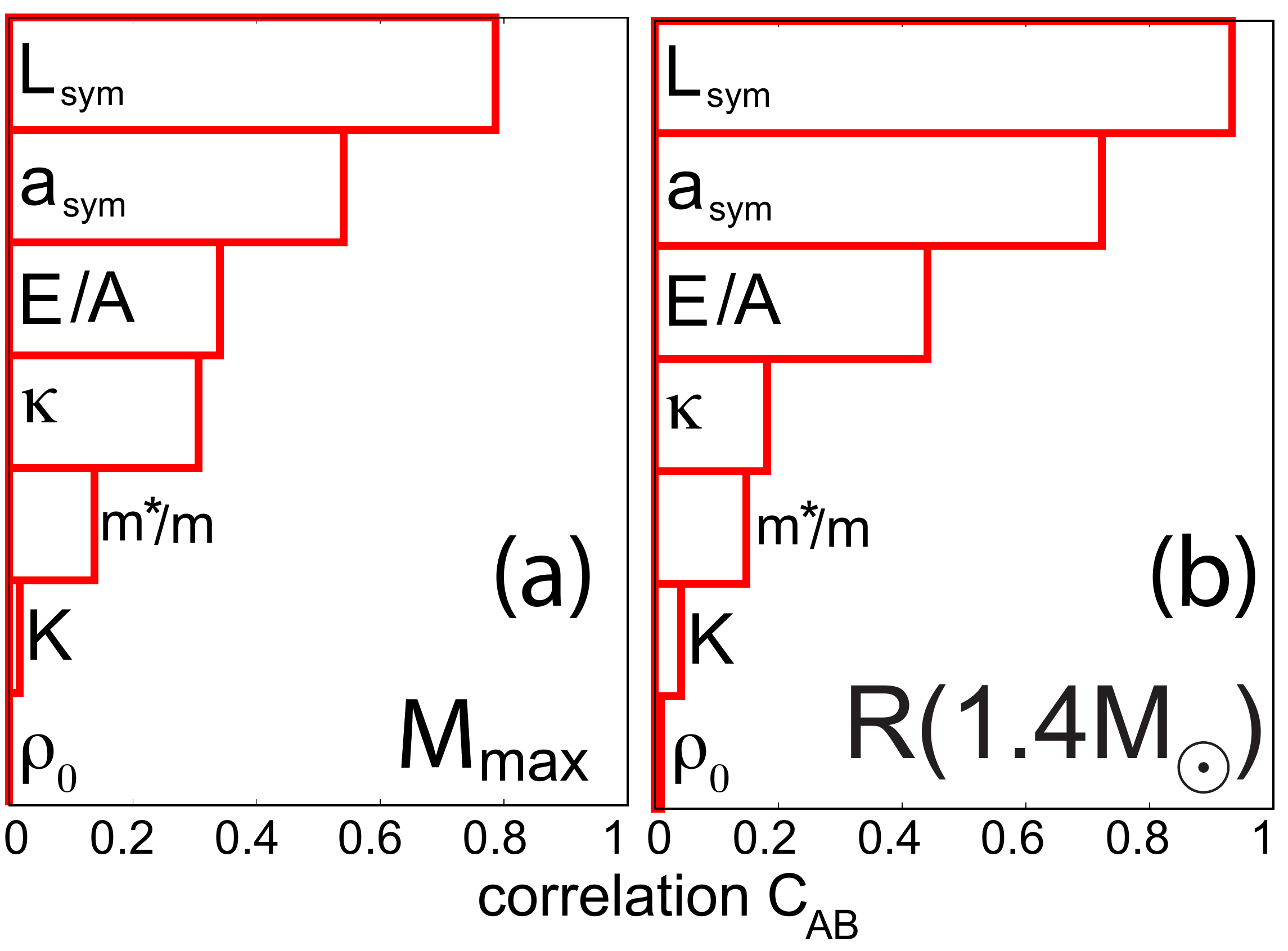}}}
\caption{\label{fig:corr-nmp-SVmin}
(Color online) Correlation $C_{AB}$ between  selected nuclear matter properties and (a) the maximum mass of a neutron star and (b) the radius of a 1.4$M_{\odot}$ neutron star obtained for the Skyrme functional SV-min.}
\end{figure}

Figure \ref{fig:corr-nmp-SVmin} presents correlation  between $M_{\rm max}$ or $R_{1.4}$ and nuclear matter properties predicted by  SV-min. 
(Note that we show  correlations for SV-min, rather than for TOV-min, because $R_{1.4}$ and $M_{\rm max}$  were explicitly constrained in optimizing  TOV-min.) 
It is seen that $R_{1.4}$ is very strongly correlated with $a_{\rm sym}$ and  $L_{\rm sym}$ while the correlation is slightly weaker for $M_{\rm max}$. Other 
nuclear matter properties, in particular the saturation density $\rho_0$, correlate weakly with the NS properties.

Table \ref{tablefour} displays the neutron skin thickness $R_{\rm skin}=r^{rms}_n - r^{rms}_p$ and electric dipole polarizability $\alpha_\mathrm{D}$ of $^{208}$Pb as predicted by  SV-min, RD-min, and TOV-min. Both observables are the focus of recent measurements and constrain the isovector part of the nuclear EDF that  is known to be poorly constrained by previous data \cite{Naz10a,Piek12,(Pie11),ReinNaz12}.  For example, the PREX experiment at the Jefferson
Laboratory has measured the neutron radius of $^{208}$Pb using parity-violating electron scattering \cite{Hor01,Abra12,PREXrapid,Horowitz:1998vv}, while the electric dipole polarizability  of $^{208}$Pb was recently accurately determined 
to be $\alpha_\mathrm{D}=14.0 \pm 0.4$\,fm$^2$/MeV  
at RCNP in Osaka in a high-resolutionn ($\vec{p},\vec{p}'$) measurement \cite{Tami11}.
The predicted values of $R_{\rm skin}$ and $\alpha_\mathrm{D}$ in
Table~\ref{tablefour} are consistent with the data  and the most recent theoretical estimates \cite{Piek12}. 
Since TOV-min yields larger value of $L_{\rm sym}$ as compared
to SV-min and RD-min, it is not surprising to see that it predicts an increased neutron skin thickness and 
$\alpha_\mathrm{D}$. Indeed, $R_{\rm skin}$  and  $\alpha_{\rm D}$
are strong isovector indicators  that are well correlated with $L_{\rm sym}$
\cite{Naz10a,Piek12}.
\begin{table}[htb]
\caption{\label{tablefour}
Neutron skin $R_{\rm skin}$ and electric dipole polarizability $\alpha_\mathrm{D}$ of  $^{208}$Pb calculated using  SV-min, RD-min and TOV-min.}
\begin{ruledtabular}
\begin{tabular}{ccc}
  \multicolumn{1}{c}{} 
  & \multicolumn{1}{c}{$R_{\rm skin}$\,(fm)}
 & \multicolumn{1}{c}{$\alpha_\mathrm{D}$\,(fm$^2$/MeV)}
\\
\hline\\[-0.2cm]
SV-min  &  0.170$\pm$0.036  &  $14.2\pm0.7$  \\
RD-min  &  0.189$\pm$0.043  &  $14.1\pm0.9$  \\
TOV-min &  0.205$\pm$0.021  &  $15.0\pm0.5$ 
\end{tabular}
\end{ruledtabular}
\end{table}

\begin{figure}[htb]
\includegraphics[width=0.8\columnwidth]{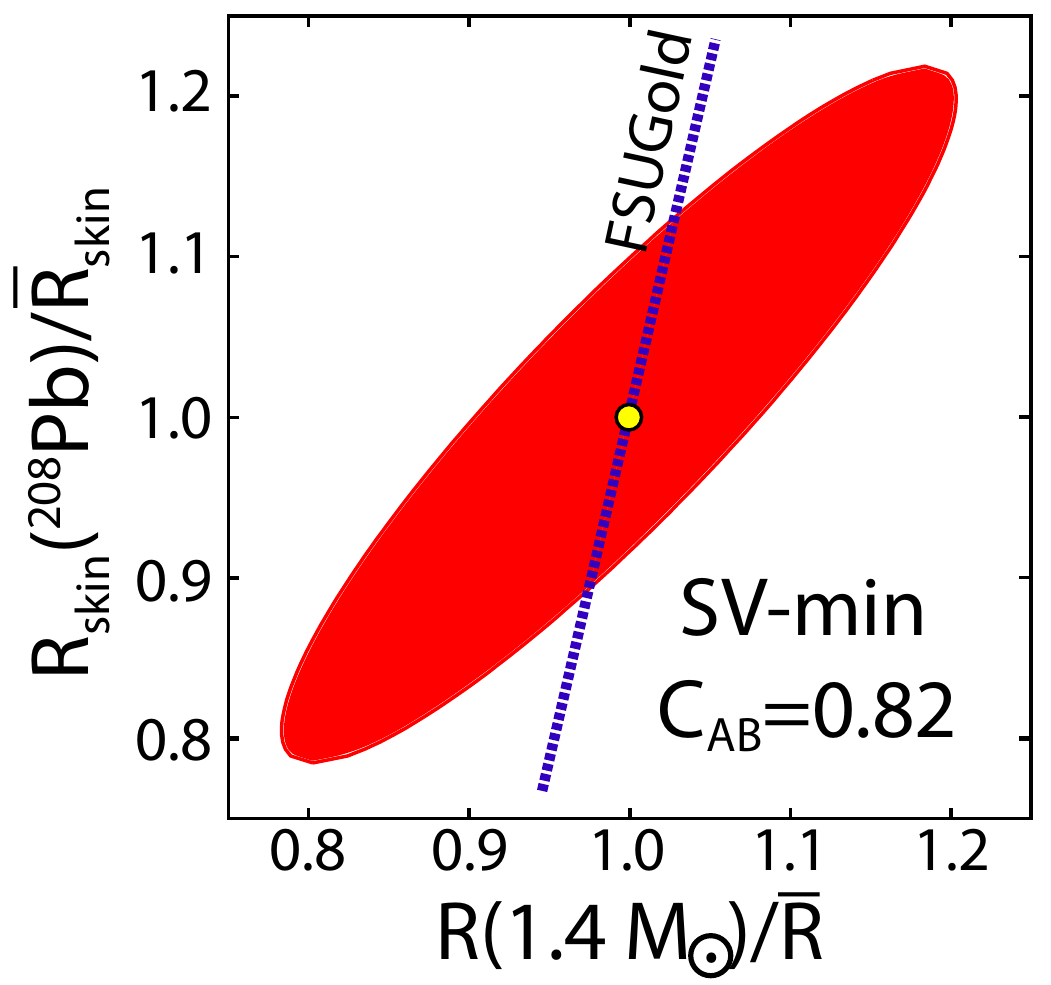}
\caption{\label{fig:corr-SVmin}
(Color online) The covariance ellipsoid for $R_{\rm skin}$ in $^{208}$Pb and the radius 
of a 1.4$M_{\odot}$ NS  calculated using SV-min. 
The mean values for SV-min are: $\bar{R}_{1.4}=10.18$\,km 
and $\bar{R}_{\rm skin}=0.170$\,fm. Also shown is the correlation line
corresponding to  FSUGold \cite{Fatt12} obtained using
${R}_{1.4}=12.66 \pm 0.46$\,km and
${R}_{\rm skin}=0.207 \pm 0.037$\,fm. 
}
\end{figure}

The covariance analysis allows one to study correlations between nuclear observables  and NS properties. Figure \ref{fig:corr-SVmin} shows the covariance ellipsoid for $R_{\rm skin}$  of $^{208}$Pb and $R_{1.4}$ calculated using SV-min.
The corresponding correlation coefficient is large, $C_{AB}=0.82$, confirming a correlation between those two observables. An even stronger correlation,
$C_{AB}=0.95$,
was obtained in Ref. \cite{Fatt12} using the relativistic EDF FSUGold.   In general, a higher pressure for neutron matter near $\rho_0$ increases both the neutron skin thickness, as neutrons are pushed out against surface tension, and $R_{1.4}$.  However, $R_{1.4}$ also depends on the pressure of neutron matter at high densities \cite{Horowitz:2001ya}.  Therefore RD-min, which has a different density dependence, predicts almost the same $R_{1.4}$ with a smaller $R_{\rm skin}$ compared to TOV-min. 
 
Note that other neutron star properties have been previously correlated with $R_{\rm skin}$, see for example \cite{Sammarruca2009,Fatt12}.  The transition density from solid crust to liquid core was found to be anticorrelated with $R_{\rm skin}$ \cite{Horowitz:2000xj,Fatt12}.  In addition, the threshold density for the rapid cooling of neutron stars via the direct URCA process was found to decrease with $R_{\rm skin}$ \cite{Horowitz:2002mb}.

\section{Applications to heavy nuclei and drip lines}
\label{sec.drip}

\begin{figure}[htb]
\includegraphics[width=\columnwidth]{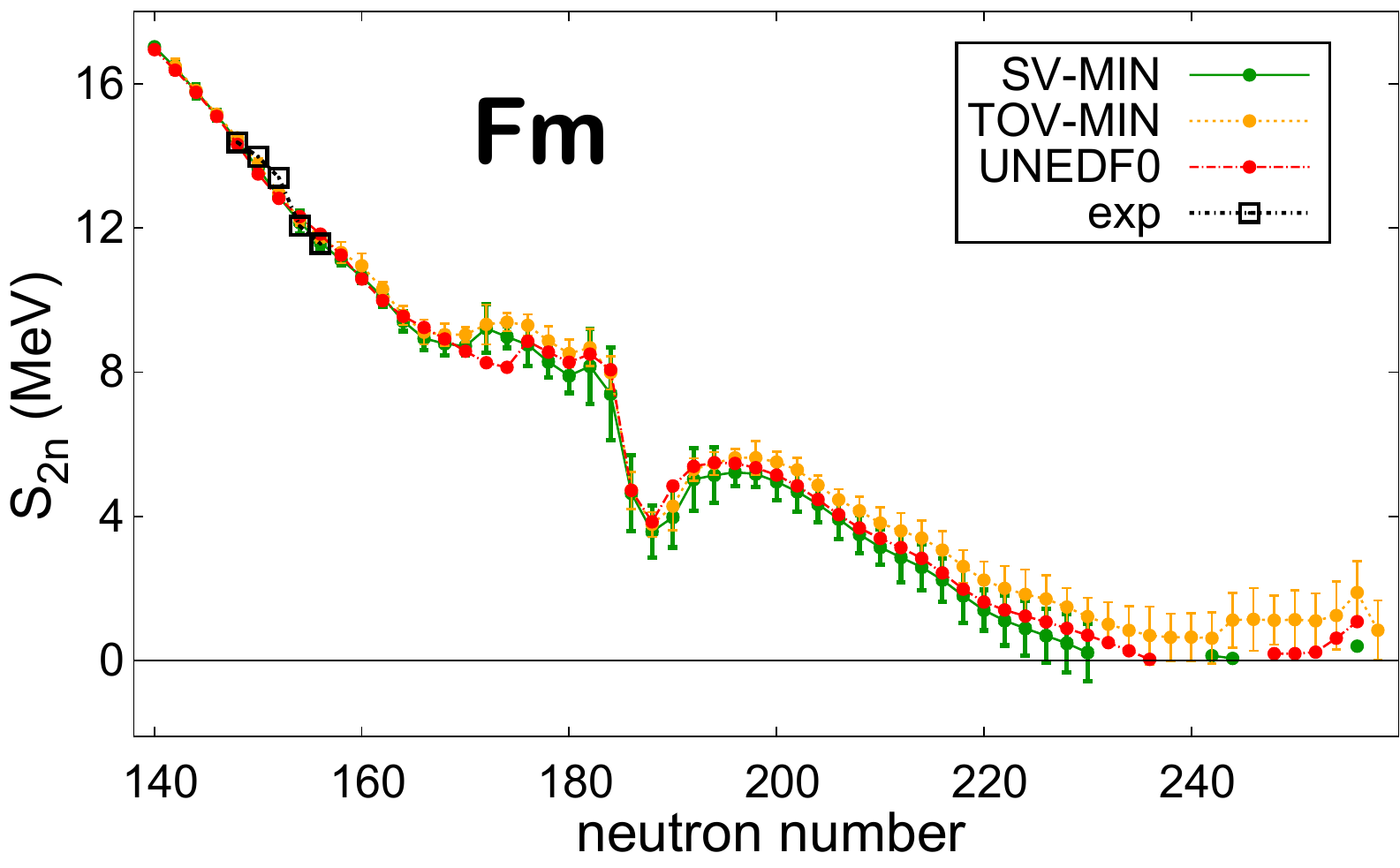}
\caption{\label{fig:S2n-Fm}
(Color online) Calculated (SV-min, UNEDF0, and 
TOV-min) and experimental two-neutron separation energies of even-even 
fermium isotopes. The error bars on SV-min and TOV-min results indicate 
statistical errors due to the uncertainty in determining their coupling constants.}
\end{figure}

\begin{figure*}[htb]
\includegraphics[width=\textwidth]{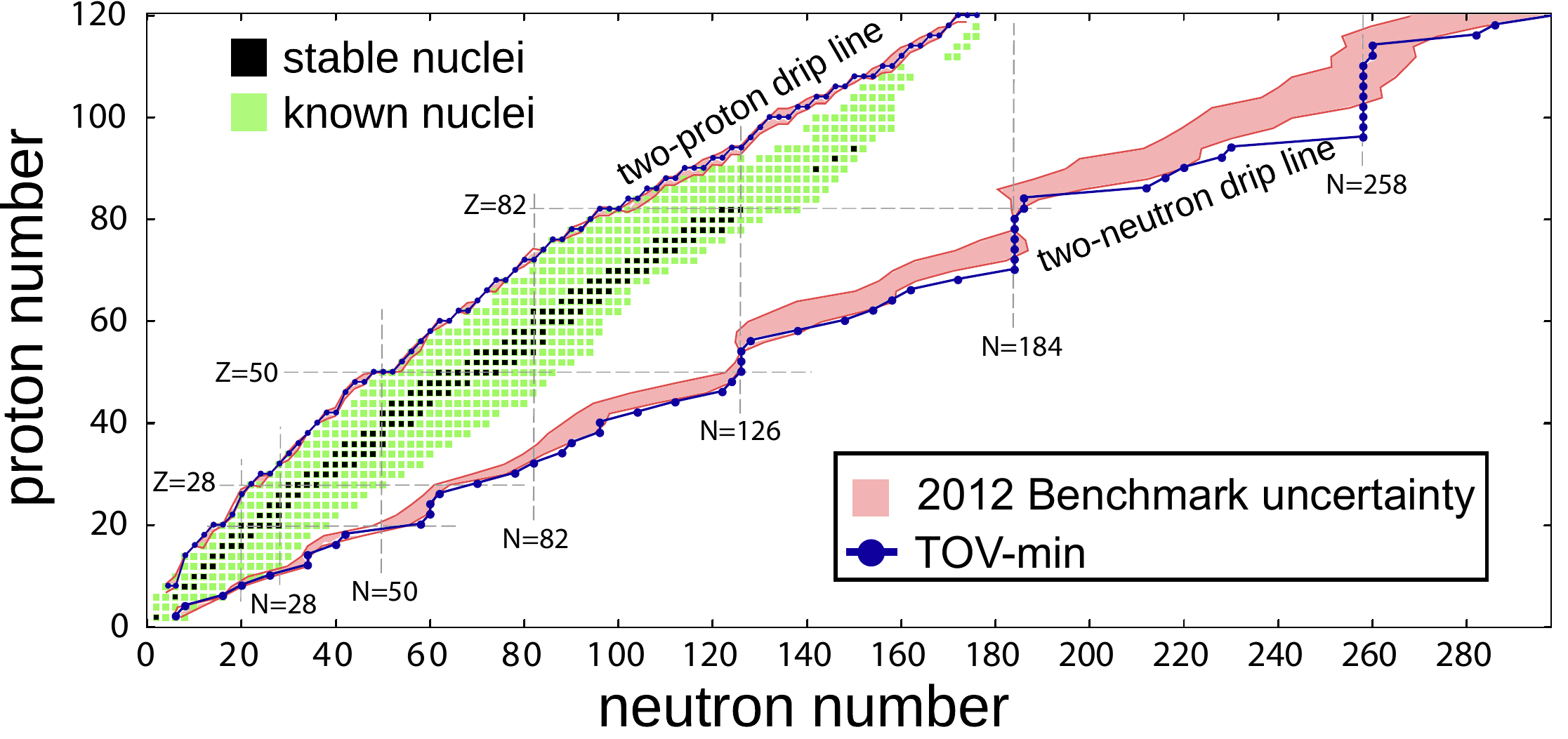}
\caption{\label{fig:drip} (Color online) Two-nucleon drip line of TOV-min (dark blue) compared to 
the uncertainty band from Ref. \cite{Erl12a} (2012 Benchmark)
obtained by averaging results of six different 
EDF: SkM*, SkP, SLy4, SV-min, UNEDF0, and UNEDF1. The dashed grey grid lines show the magic numbers 
known around the valley of stability (20, 28, 50, 82, 126) as well as the predicted regions of enhanced
shell-stability in superheavy nuclei around $N=184$ and 258.  All 767 experimentally known even-even isotopes are shown with the stable nuclei indicated as black squares and the radioactive nuclei as green squares.
}
\end{figure*}

As seen in Sec.~\ref{cha:corr}, adding NS  data helps constrain isovector nuclear matter properties and the properties of neutron-rich nuclei such as the neutron skin  thickness. These additional data should allow one to make more accurate predictions for very neutron-rich nuclei.  In this section, we study the effect of NS data on the structure of nuclei near the drip lines.  Figure \ref{fig:S2n-Fm} shows two-neutron separation energies of even-even fermium isotopes using SV-min, UNEDF0, and TOV-min.  For isotopes where experimental data is available, the models agree and nicely reproduce the experimental data as discussed in \cite{Erl12a}. 
Moving to more neutron rich isotopes the discrepancy between various predictions grows and ends up in different predictions for the location of the neutron drip line (S$_{2n}$=0).  In the same way, statistical error bars for SV-min and TOV-min grow with increasing neutron number.  However, TOV-min shows smaller error bars compared to SV-min because of the additional constraints on NS  properties. Indeed the theoretical uncertainty band
for the mass-radius relation is dramatically reduced when going from  SV-min to TOV-min, see  Fig.~\ref{fig:uncert-SVmin}. 

The  functionals SV-min and UNEDF0 show the phenomenon of re-entrant binding caused by shell effects \cite{Erl12a} where some heavier isotopes are bound beyond the first two-neutron drip line.  However the separation energies for TOV-min are slightly larger and  do not exhibit  the re-entrant behavior.

A global survey of the two-nucleon drip lines  predicted with TOV-min up to proton number $Z=120$ is summarized in Fig. \ref{fig:drip}.  The TOV-min  results are compared to the uncertainty band of a systematic study \cite{Erl12a}  (2012 Benchmark) using a variety of different nuclear EDFs.    Compared to the uncertainty band, the two-neutron drip line for TOV-min is slightly shifted towards more neutron-rich nuclei, and it lies near the outer border of the error band of 2012 Benchmark.    A reason for deviations of TOV-min from the error band around shell closures at $N=184$ and $N=258$ could be the non-existence of re-entrant binding for TOV-min as seen in Fig.~\ref{fig:S2n-Fm} due to an increased binding (and $S_{2n}$) of neutron-rich nuclei.   For the two-proton drip line, the TOV-min results agree very well with  2012 Benchmark.

\section{Summary and Conclusions}
\label{sec.conclusions}

In this study, we proposed and investigated new approaches for predictions of NS properties based on Skyrme functionals. We have shown that standard  Skyrme functionals, adjusted to properties of finite nuclei,  can not reproduce the expected maximum neutron star mass $M_{\rm max}$ and the radius of a 1.4$M_\odot$ neutron star $R_{1.4}$, and that predictions can only be made within an extremely broad uncertainty band.

In a first step, we calculated a new EOS by matching a high-density EDF, that gives reasonable NS properties, to a low-density EDF adjusted to nuclear properties.  Using this approach, by matching  LS220 to UNEDF0,
it is possible to predict a maximum mass of about 2.2$M_\odot$, but the radius of a 1.4$M_\odot$ neutron star is relatively small, ranging from 10 to 11 km. 
Based on this exercise, we can  conclude that the low-density part of the EOS carries no information on $M_{\rm max}$. Therefore, EDFs optimized to nuclear data cannot be used to predict this quantity, and scrutinizing existing functionals with respect to $M_{\rm max}$ makes little sense.

In a next step,  we have optimized a new Skyrme functional TOV-min by simultaneously considering nuclear and NS data.  Results for standard nuclear bulk properties  are as good as results obtained with the established  functionals. Nuclear matter properties vary slightly for isoscalar properties between SV-min and TOV-min while isovector properties are considerably changed.
This is not surprising because the NS data mainly constrain the isovector channel. In particular, 
the values of $a_{\rm surf}$ and $L_{\mathrm{sym}}$ of TOV-min are increased 
with respect to SV-min, but they are still consistent  with the experimentally allowed  range \cite{TsangRev, LattRev}. Interestingly, the RD-min functional that has a modified density dependence does also very well on $M_{\rm max}$ and $R_{1.4}$, in spite of the fact that it has not been optimized to NS data; its values of $a_{\rm surf}$ and $L_{\mathrm{sym}}$ are fairly close to those of TOV-min.

A quantity that contains  information on  the isovector channel is the neutron skin thickness.  Adding NS data leads to an increased neutron skin for TOV-min compared to SV-min.  Furthermore, a previously reported correlation between neutron skin and the NS radius is confirmed.
We further checked the predictions of TOV-min for neutron rich nuclei and for the position of two-particle drip lines. Results for separation energies based on the new functional agree with established Skyrme functionals, such as SV-min or UNEDF0, in the regions where experimental data are available.  
The two-neutron drip line predicted by  TOV-min is shifted towards the more neutron-rich side  of  the 2012 Benchmark uncertainty band   \cite{Erl12a}, while the two-proton drip line of TOV-min lies within  the 2012 Benchmark.

The new functional TOV-min is a first example of EDF optimization using  both nuclear and neutron star data.  While the results obtained with TOV-min are very encouraging, various improvements are anticipated in the near future. Those include the use of a large nuclear database containing deformed nuclear states,  advanced optimization protocol as in Refs.~\cite{UNEDF0,UNEDF1}, and  improved description of neutron star radius data by considering more realistic crust models.  This work will be carried out under the Nuclear Low Energy Computational Initiative (NUCLEI) \cite{NUCLEI}.

In summary, we have successfully optimized  a new EDF that is applicable for both finite nuclei and neutron stars.  NS data is used to better constrain isovector interactions.  As a result, the statistical extrapolation errors on predicted observables in neutron-rich nuclei are reduced with TOV-min; hence,  the functional is expected to yield more reliable predictions in the region of  very neutron-rich heavy nuclei.

%
\begin{acknowledgments}
This work was supported  by the Office of
Nuclear Physics, U.S. Department of Energy under Contract Nos.
DE-FG02-96ER40963 (University of Tennessee), DE-SC0008499 and DE-SC0008808    (NUCLEI SciDAC Collaboration),  DE-FG02-87ER40365 (Indiana University), and by BMBF under contract No. 06~ER~9063.
\end{acknowledgments}


\bibliographystyle{apsrev4-1}
\bibliography{NS}

\end{document}